\documentclass[aps,prl,showpacs,groupedaddress,twocolumn]{revtex4}

\usepackage{amsmath}
\usepackage{amssymb}
\usepackage{bm}
\usepackage{graphicx}
\usepackage{color}
\usepackage{pdfpages}

\begin{document}

\title{Symmetry and Size of Membrane Protein Polyhedral Nanoparticles}

\author{Di Li}
\author{Osman Kahraman}
\author{Christoph A. Haselwandter}
\affiliation{Department of Physics \& Astronomy and Molecular and Computational Biology Program, Department of Biological Sciences, University of Southern California, Los Angeles, CA 90089, USA}

\begin{abstract}

In recent experiments [T. Basta \textit{et al.}, Proc. Natl. Acad. Sci. U.S.A. \textbf{111}, 670 (2014)] lipids and membrane proteins were observed to self-assemble into membrane protein polyhedral nanoparticles (MPPNs) with a well-defined polyhedral protein arrangement and characteristic size. We develop a model of MPPN self-assembly in which the preferred symmetry and size of MPPNs emerge from the interplay of protein-induced lipid bilayer deformations, topological defects in protein packing, and thermal effects. With all model parameters determined directly from experiments, our model correctly predicts the observed symmetry and size of MPPNs. Our model suggests how key lipid and protein properties can be modified to produce a range of MPPN symmetries and sizes in experiments.

\end{abstract}


\maketitle

\begin{figure}[b!]
\centerline{\includegraphics[scale=0.24]{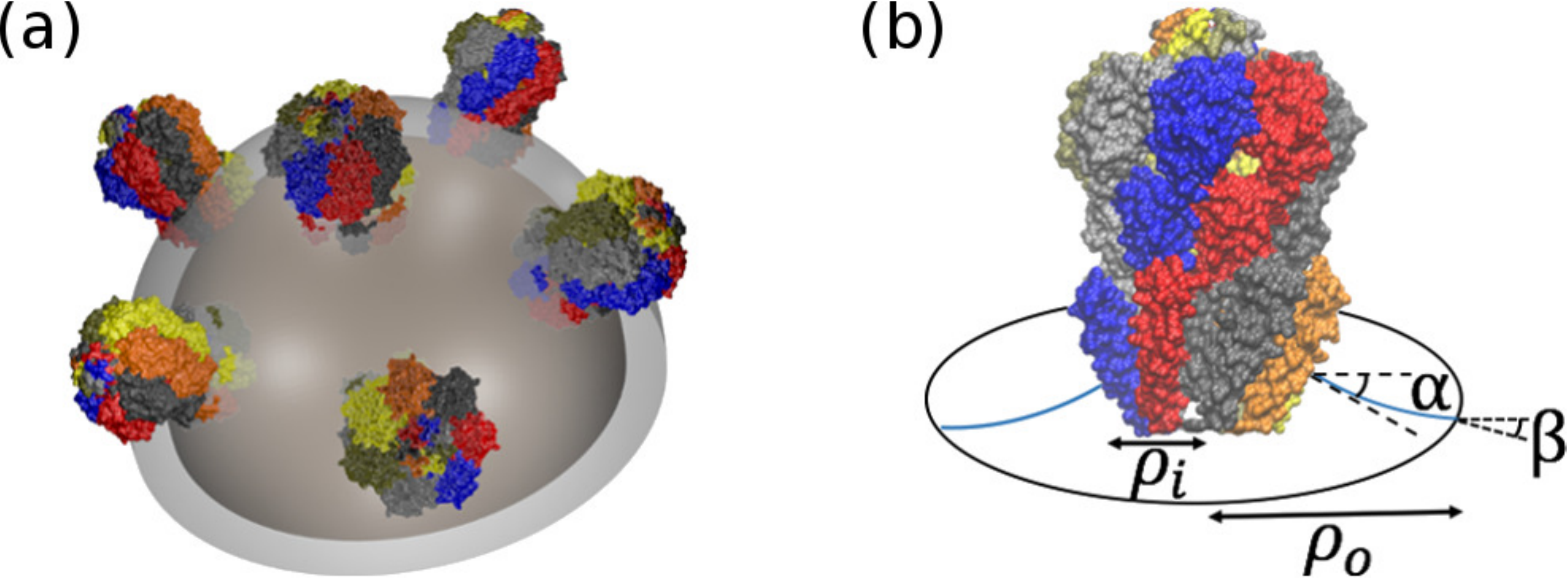}}
\caption{\label{fig:one} (color online). Schematic of MPPNs \cite{Basta2014}. (a) MscS are embedded in a lipid bilayer with the MscS cytoplasmic region outside MPPNs \cite{Basta2014}. (b) Lipid bilayer bending deformations (blue curves) induced by the observed MscS structure \cite{Bass2002,Steinbacher2007,Phillips2009} with protein radius $\rho_i$ and bilayer-protein contact angle $\alpha$. The membrane patch radius $\rho_o$ and boundary angle $\beta$ are determined by the MPPN size and the number of proteins per MPPN. (MscS Protein Data Bank ID 2OAU with \cite{VMD1996} different colors indicating different MscS subunits.)}
\end{figure}

Membrane proteins play a central role in a variety of essential cellular processes \cite{McMahon2005,Engelman2005} such as ion exchange, signaling, and membrane curvature regulation. The biologically relevant structures, and resulting functions, of many membrane proteins depend critically \cite{Engelman2005,Phillips2009} on their lipid bilayer environment, and on chemical and voltage gradients across the cell membrane. Yet, determination of membrane protein structure in lipid bilayer environments and in the presence of physiologically relevant transmembrane gradients has largely remained elusive. Recent experiments on membrane protein polyhedral nanoparticles (MPPNs) \cite{Basta2014} present an exciting step towards overcoming this challenge. In these experiments, lipids and mechanosensitive channels of small conductance (MscS) \cite{Bass2002,Steinbacher2007} were observed to self-assemble \cite{Basta2014,Wu2013} into lipid bilayer vesicles with a polyhedral protein arrangement [see Fig.~\ref{fig:one}(a)]. The dominant polyhedral symmetry of MPPNs was found \cite{Basta2014} to be the snub cube (snub cuboctahedron) with MscS located at its 24 vertices and a characteristic MPPN radius $\approx20$~nm. The well-defined symmetry and characteristic size of MPPNs may permit \cite{Basta2014} structural analysis of membrane proteins with the membrane proteins embedded in a lipid bilayer environment and the closed surfaces of MPPNs supporting physiologically relevant transmembrane gradients.

Utilization of MPPNs for high-resolution structural studies requires \cite{Basta2014}
control over the symmetry and size of MPPNs. In this Letter we develop a physical description of MPPNs which establishes a quantitative link between the shape of MPPNs and key molecular properties of their constituents. We first describe a simple mean-field model of MPPNs inspired by previous work on membrane budding \cite{Gozdz2001,Auth2009,Muller2010} and viral capsid self-assembly \cite{Bruinsma2003}. Our mean-field model of MPPNs accounts for the lipid bilayer bending deformations induced by MscS \cite{Bass2002,Steinbacher2007,Phillips2009} and the MscS packing defects resulting from the spherical topology of MPPNs, and yields the MPPN energy as a function of the number of MscS per MPPN without any free parameters. We confirm some of the key assumptions underlying our mean-field model by carrying out Monte Carlo simulations of a minimal molecular model of MPPN organization, which we formulate following previous work on viral capsid symmetry \cite{Zandi2004}. Finally, we use our mean-field model of MPPNs to calculate \cite{Bruinsma2003,Safranbook,BenShaul1994} the MPPN self-assembly phase diagram as a function of protein concentration, bilayer-protein contact angle, and protein size. We show that our model correctly predicts, with all model parameters determined directly from experiments, the observed \cite{Basta2014} symmetry and size of MPPNs formed from MscS. Our results suggest that the preferred symmetry and size of MPPNs emerge from the interplay of protein-induced lipid bilayer deformations, topological defects in protein packing, and thermal effects.

\textit{Mean-field model.}\textbf{---}The membrane-spanning region of MscS \cite{Bass2002,Steinbacher2007} has an approximately conical shape \cite{Phillips2009,sm} with radius $\rho_i\approx 3.2$~nm in the lipid bilayer midplane and bilayer-protein contact angle $\alpha \approx 0.46$--$0.54$~rad, yielding \cite{Phillips2009} protein-induced lipid bilayer bending deformations [see Fig.~\ref{fig:one}(b)]. The preferred MscS arrangement minimizing bilayer bending energy is expected \cite{Gozdz2001,Auth2009,Muller2010} to be a uniform hexagonal lattice. Our simple mean-field model of MPPNs therefore considers, on the one hand, contributions to the MPPN energy arising from MscS-induced bilayer bending deformations for hexagonal protein arrangements \cite{Dan1994,Gozdz2001,Auth2009,Muller2010}. On the other hand, the spherical shape of MPPNs necessitates defects in the preferred hexagonal packing of MscS which, in analogy to viral capsids \cite{Bruinsma2003,Zandi2004}, yields an energy penalty characteristic of the number of proteins per MPPN, $n$. Thus, we allow in the total MPPN energy $E=E_b+E_d$ for contributions due to protein-induced bilayer bending, $E_b$, and topological defects in protein packing, $E_d$, respectively.

We estimate the MPPN bending energy $E_b$ using a formalism developed in the context of membrane budding \cite{Gozdz2001,Auth2009,Muller2010}, which we summarize here for completeness. The unit cell associated with uniform
hexagonal protein arrangements can be approximated \cite{Gozdz2001,Auth2009,Muller2010} by a circular membrane patch with the protein at its center and boundary conditions set by the spherical shape of MPPNs. We denote the projected radius of the circular membrane patch by $\rho_o=R \sin\beta$ [Fig.~\ref{fig:one}(b)], where $R$ is the bilayer midplane radius of MPPNs and $\beta$ is the patch boundary angle. We have $\beta =\arccos[(n-2)/n]$ so that the total area covered by membrane patches, $2 n \pi R^2 (1-\cos \beta)$, is equal to the total MPPN area, $4 \pi R^2$. Minimization of the Helfrich-Canham-Evans bending energy \cite{Canham1970,Evans1974,Helfrich1973,Boal2012} with respect to the bilayer midplane height field then yields \cite{Auth2009} the MPPN bending energy,
\begin{equation} \label{eq:two}
E_b(n,R)=\frac{2n\pi K_b \left(b\rho_o-a\rho_i\right)^2}{\rho_o^2-\rho_i^2}\,,
\end{equation}
where the bilayer bending rigidity $K_b \approx 14$~$k_B T$ \cite{Rawicz2000}
for the diC14:0 lipids used for MPPNs formed from MscS \cite{Basta2014,Wu2013}, and the slopes at the bilayer-protein boundary $a=-\tan \alpha$ and at the
membrane patch boundary $b=-\tan \beta$. To account for steric constraints on lipid and protein size we only allow \cite{sm} in Eq.~(\ref{eq:two}) for membrane patch sizes $\geqslant \rho_i+\rho_l$ when calculating the MPPN energy, where the lipid radius $\rho_l \approx 0.45$~nm for diC14:0 lipids \cite{Damodaran}. Equation~(\ref{eq:two}) yields \cite{Auth2009} a preferred unit cell size of hexagonally packed proteins, with $E_b=0$, which can be
achieved if $\alpha>\beta$, corresponding to close-packed catenoidal bilayer deformation profiles.

\begin{figure}[t!]
\centerline{\includegraphics[scale=0.28]{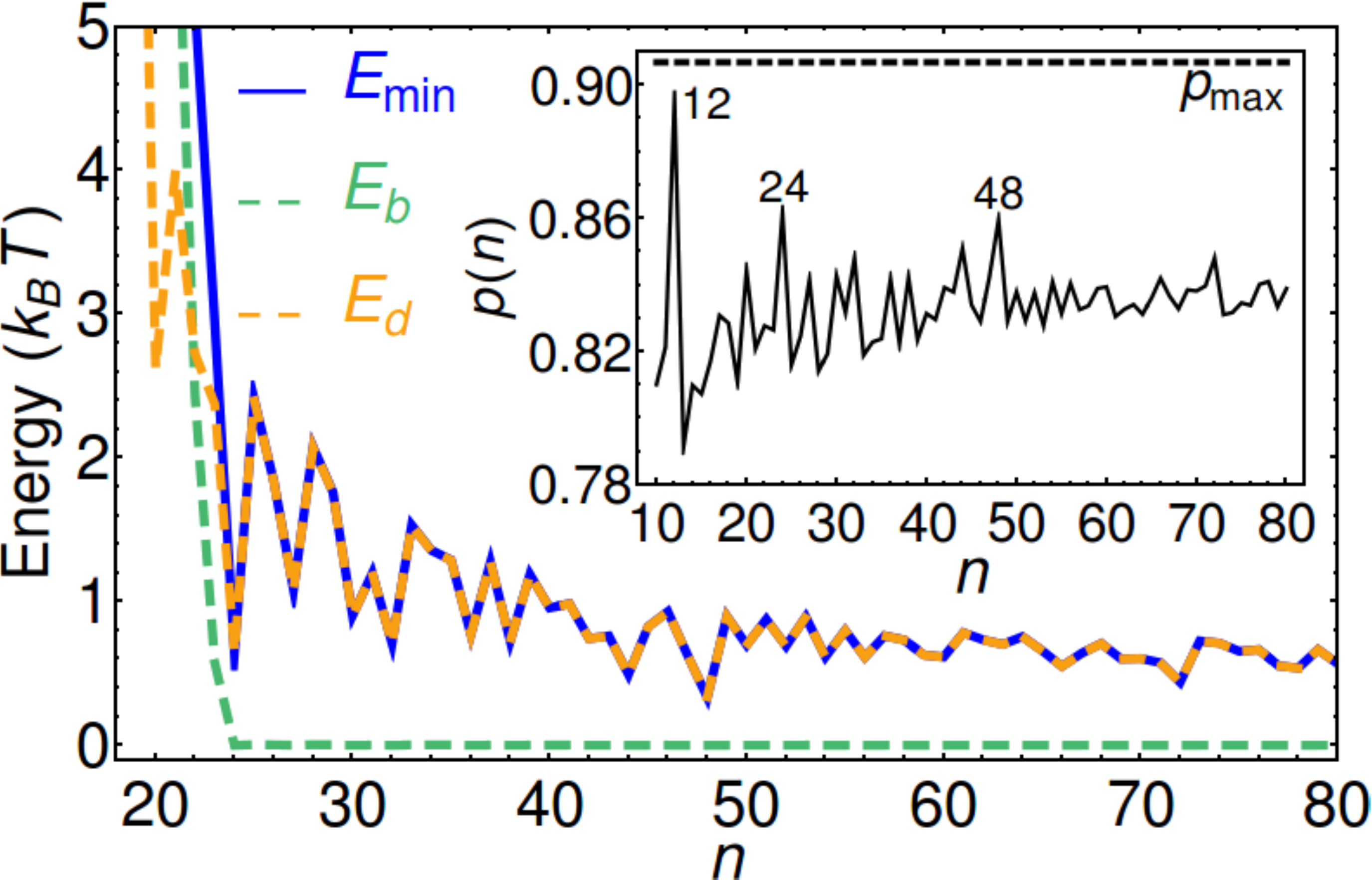}}
\caption{\label{fig:two} (color online). Minimized total MPPN energy $E_{\text{min}}$
for MPPNs formed from MscS \cite{Basta2014,Wu2013} with the contributions $E_b$ and $E_d$ due to bending deformations and packing defects versus $n$ at $\alpha=0.5$~rad. The inset shows the optimal sphere coverage $p(n)$ for $n$ identical non-overlapping circles \cite{Clare1991} with particularly
favorable packings at $n=12$ (icosahedron), $n=24$ (snub cube), and $n=48$.
(Inset after Ref.~\cite{Clare1991}.)}
\end{figure}

Topological defects perturb the hexagonal packing of proteins in MPPNs. To estimate, for each $n$ and $R$, the energy cost of the resulting deviations from the preferred protein arrangement in Eq.~(\ref{eq:two}) we adopt a mean-field approach developed in the context of viral capsid self-assembly \cite{Bruinsma2003} and approximate the hexagonal spring network associated with Eq.~(\ref{eq:two}) \cite{Auth2009} by a uniform elastic sheet \cite{Phillips2012}. In the continuum limit, the minimum-energy protein arrangement in Eq.~(\ref{eq:two}) satisfying $\rho_o >\rho_i$ then yields \cite{Kantor1987,sm} the stretching modulus
\begin{equation} \label{eq:threeKs}
K_s=\frac{\pi K_b}{2\sqrt{3}}\frac{\text{min}\left(a^4,b^4\right)}{\left|a^2-b^2\right|\rho_i^2}\,,
\end{equation}
which depends on $n$ via $b$. At the mean-field level, the deviation from the preferred hexagonal packing of proteins in Eq.~(\ref{eq:two}) due to the spherical shape of MPPNs can be quantified \cite{Bruinsma2003} through the fraction of the surface of a sphere enclosed by $n$ identical non-overlapping circles at closest packing \cite{Clare1991}, $p(n)$, which yields
in our model the protein arrangement at each $n$. The local maxima of $p(n)$ correspond to locally optimal protein packings [see Fig.~\ref{fig:two}(inset)]. We thus find \cite{sm} the MPPN defect energy
\begin{equation}
E_d(n,R)=2 \pi K_s R^2 \left[\frac{p_{\text{max}}-p(n)}{p_{\text{max}}}\right]^2\,,
 \label{eq:six}
\end{equation}
where $p_{\text{max}}=\pi/2\sqrt{3}$ corresponds to the uniform hexagonal protein arrangements assumed in Eq.~(\ref{eq:two}).

We minimize the total MPPN energy $E$ given by Eqs.~(\ref{eq:two})--(\ref{eq:six}), at each $n$, with respect to $R$, which yields the minimum MPPN energy $E_{\text{min}}(n)$
with all parameters determined directly by the molecular properties of the lipids and proteins forming MPPNs (see Fig.~\ref{fig:two}). We find that
MPPNs with $n<n_0$, where $n_0\approx20$ for MPPNs formed from
MscS \cite{Basta2014,Wu2013} with $\alpha\approx 0.5$~rad, are strongly penalized by the MPPN bending energy, which cannot be minimized to zero in this regime. Furthermore, MPPNs with $n<n_0$ also tend to be penalized by the MPPN defect energy because $K_s$ in Eq.~(\ref{eq:threeKs}) can be large for small $n$ \cite{sm}. For $n>n_0$, in which case also $\alpha>\beta$, we find a range of favorable $n$ corresponding to locally optimal protein packings. However, for $n>n_0$ the MPPN energies associated with distinct $n$ fall within just a few $k_B T$ of each other and, as we discuss further below, thermal effects are therefore crucial in this regime. Finally we note that, for $n$ which allow $E_b=0$ in Fig.~\ref{fig:two}, the preferred protein separation (and, hence, MPPN size) is set, within $0.5\%$, by $E_b$ in Eq.~(\ref{eq:two}).

\textit{Minimal molecular model.}\textbf{---}Our mean-field model of MPPNs assumes \cite{Gozdz2001,Auth2009,Muller2010,Bruinsma2003} that, for a given $n$, the protein arrangement in MPPNs is determined by close packing of circular membrane patches, each with a protein at its center. Following previous work on viral capsid symmetry \cite{Zandi2004}, we test these assumptions through Monte Carlo simulations of a minimal molecular (particle-based) model of MPPN organization, which focuses on short-range interactions between lipids
and proteins. In this model, we represent \cite{sm} the lipid bilayer and membrane proteins by disks lying on the surface of a sphere and assume that lipids interact with other lipids and proteins via Lennard-Jones potentials~\cite{Zandi2004} with, for simplicity, hardcore steric repulsion between proteins. These interactions can be parametrized \cite{sm} based on experiments and previous calculations \cite{Steinbacher2007,Damodaran,Boal2012,BenTal1996,Choe2008} but our simulation results are not sensitive to the particular interactions used. We employed simulated annealing Monte Carlo simulations \cite{Kirkpatrick1983} with linear cooling to numerically determine the minimum-energy configuration of lipids and proteins in our minimal molecular model of MPPN organization. Following experiments on MPPNs formed from MscS \cite{Basta2014} we focused in our simulations on MPPNs with $n=24$ and a total of $\approx1700$ lipids~\cite{sm}.

\begin{figure}[t!]
\centerline{\includegraphics[width=\columnwidth]{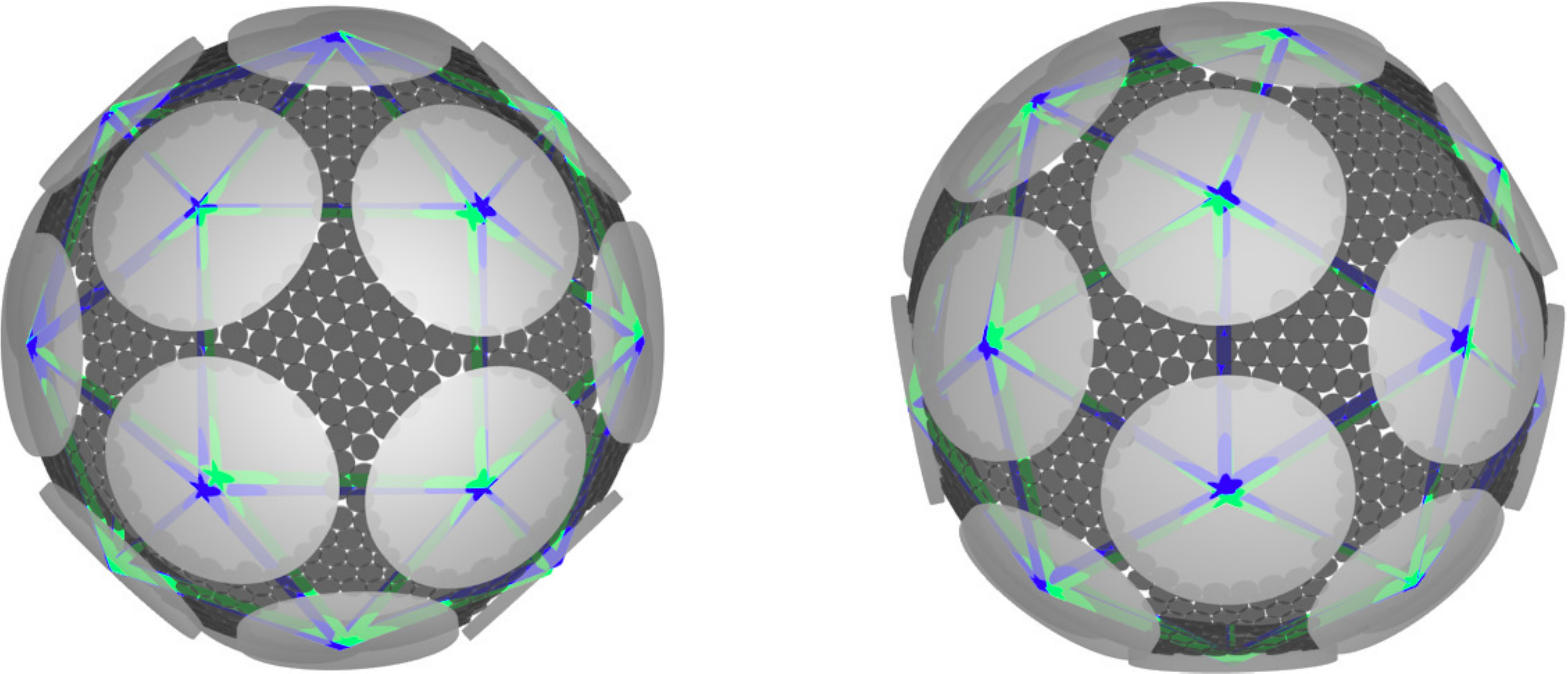}}
\caption{\label{fig:three} (color online). Front and side views of the minimum-energy MPPN configuration obtained from our minimal mo\-lecular model of MPPN organization. The larger and smaller disks represent proteins and the lipid bilayer, with disk sizes corresponding to \cite{sm} MscS \cite{Steinbacher2007} and diC14:0 lipids \cite{Damodaran}, respectively. The green and blue lines are obtained by connecting the centers of nearby MscS in the simulated MPPN configuration and by fitting the simulated MPPN configuration to a snub cube (dextro) using least-square minimization.}
\end{figure}

Figure~\ref{fig:three} shows the minimum-energy MPPN configuration found in our simulations. The results in Fig.~\ref{fig:three} suggest that, in the ground state of the system, MscS are arranged in the form of a snub cube, in agreement with the corresponding optimal protein packing assumed in the mean-field model of MPPNs in Fig.~\ref{fig:two}. To quantify the quality of the polyhedral fit in Fig.~\ref{fig:three} we proceeded as in experiments on MPPNs \cite{Basta2014} and used least-square minimization \cite{sm} to calculate the minimum fit error for 132 convex polyhedra \cite{Encyclopedia}: the Platonic, Archimedean, Catalan, and Johnson solids. We define \cite{Basta2014} the fit error as the sum over the squared distances between the simulated positions of protein centers and the closest fitted polyhedron vertices. We find \cite{sm} that the snub cube (dextro) yields the best fit with a fit error $\approx 420$~\AA$^2$, while the second- and third-best fits are provided by the truncated cuboctahedron and pentagonal hexecontahedron (levo) with the substantially larger fit errors $\approx 3700$~\AA$^2$ and $\approx 5800$~\AA$^2$, respectively.

\textit{Phase diagram.}\textbf{---}Based on our mean-field model of MPPNs
we construct the MPPN phase diagram from \cite{Bruinsma2003} the statistical thermodynamics of amphiphile self-assembly in dilute aqueous solutions \cite{Safranbook,BenShaul1994}. Let $N_n$ denote the total number of proteins bound in MPPNs with $n$ proteins each and $N_w$ the total number of solvent molecules, which we take \cite{Basta2014,Wu2013} to be dominated by contributions due to water. For MPPNs formed from MscS
the protein concentration $\approx 1$~mg/mL \cite{Basta2014}, with the molecular mass $\approx 2.2\times10^{5}$~g/mol for MscS \cite{Vasquez2007}, yielding the protein number fraction $c=\sum_n N_n/N_w\approx7.8\times10^{-8}$. We assume here that all proteins in the system are incorporated into MPPNs, a point we return to below. In the dilute limit $c \ll 1$ with no interactions between MPPNs we have \cite{Safranbook,BenShaul1994} the mixing entropy $S=-N_w k_B\sum_n \Phi(n) [\text{ln}\Phi(n)-1]$, where the MPPN number fraction $\Phi(n)=\frac{N_n}{n N_w}$. This then allows \cite{Safranbook,BenShaul1994} construction of the Helmholtz free energy $F=E-T S$ with $E=N_w \sum_n \Phi(n) E_{\text{min}}(n)$, in which the minimum MPPN energy $E_{\text{min}}(n)$
is determined by our mean-field model of MPPNs via Eqs.~(\ref{eq:two})--(\ref{eq:six}). Minimization of $F$ with respect to $\Phi(n)$ \cite{Safranbook,BenShaul1994}
results in
\begin{equation}
\Phi(n)=e^{\beta[\mu n-E_{\text{min}}(n)]}\ ,
\label{eq:nine}
\end{equation}
where $\beta=1/k_B T$ and the protein chemical potential $\mu$ is fixed by
the constraint $\sum_n n \Phi(n)=c$. As in Fig.~\ref{fig:two}, we restrict $n$ to the range $10 \leq n \leq 80$ for simplicity. Finally, we calculate the MPPN equilibrium distribution $\phi(n)$ from Eq.~(\ref{eq:nine}) via $\phi(n)=\Phi(n)/\sum_{n=10}^{80} \Phi(n)$.

\begin{figure}[t!]
\centerline{\includegraphics[scale=0.22]{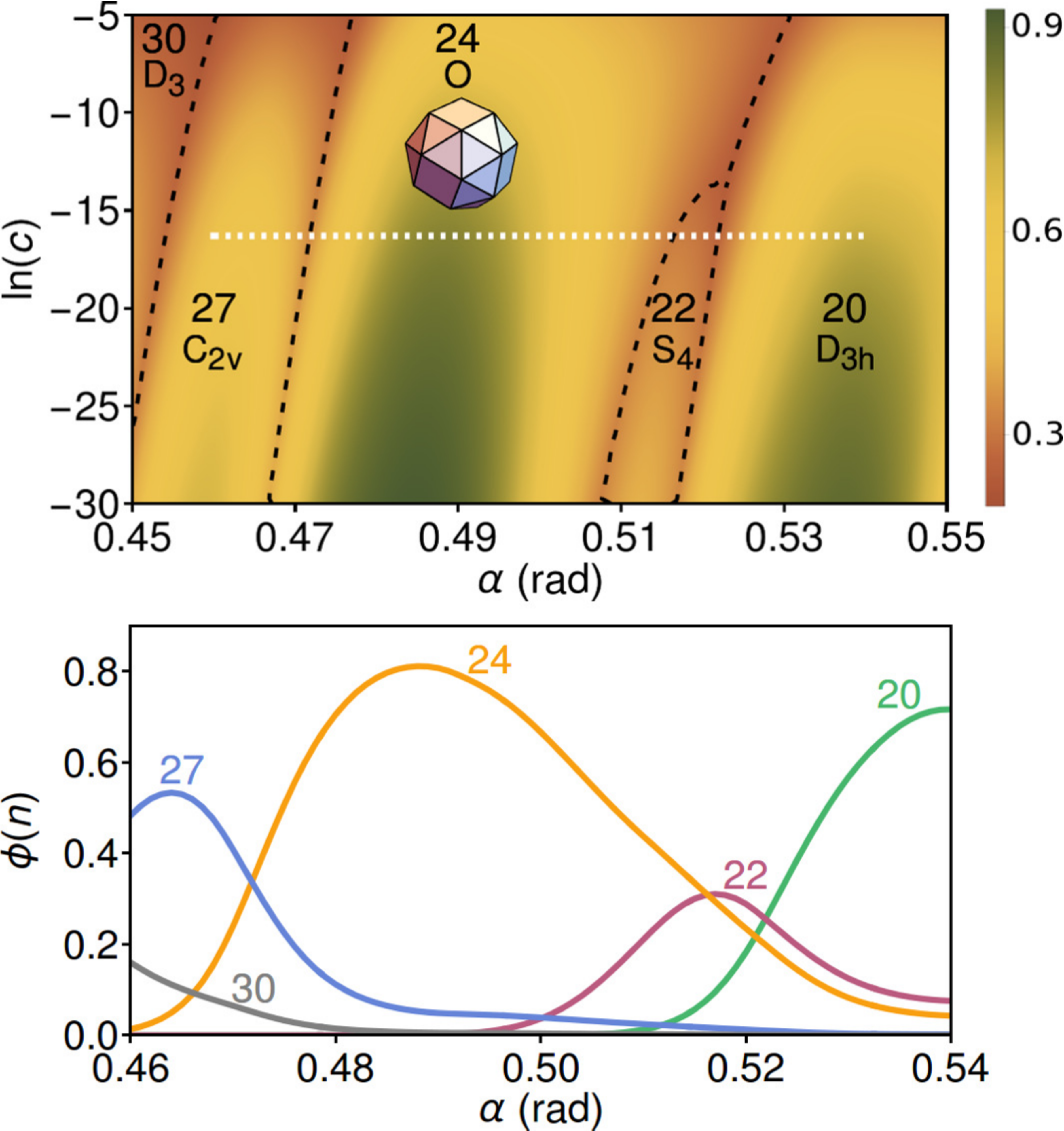}}
\caption{\label{fig:four} (color online). MPPN self-assembly phase diagram
obtained from Eq.~(\ref{eq:nine}) as a function of protein number fraction $c$ and bilayer-protein contact angle $\alpha$. The color map in the upper panel shows the maximum values of $\phi(n)$ associated with the dominant $n$-states of MPPNs. The dominant $n$ are indicated in each portion of the phase diagram, together with the associated MPPN symmetry \cite{Clare1991}. Black dashed curves delineate phase boundaries. The white dashed line indicates the value $c\approx7.8\times10^{-8}$ corresponding to experiments on MPPNs \cite{Basta2014} and the $\alpha$-range associated with MscS \cite{Steinbacher2007,Phillips2009,sm}. The lower panel shows $\phi(n)$ for $n=20$, $22$, $24$, $27$, and $30$ as a function of $\alpha$ along the white dashed line in the upper panel. We use the model parameter values $\rho_i\approx 3.2$~nm \cite{Steinbacher2007,Phillips2009,sm} and $K_b \approx 14$~$k_B T$ \cite{Rawicz2000} corresponding to MPPNs formed from \cite{Basta2014,Wu2013} MscS and diC14:0 lipids.}
\end{figure}

Figure~\ref{fig:four} shows the MPPN self-assembly phase diagram as a function of protein number fraction $c$ and bilayer-protein contact angle $\alpha$
for the region of parameter space relevant for MPPNs formed from MscS \cite{Basta2014,Wu2013}. In agreement with experiments \cite{Basta2014,Wu2013} our model predicts that MPPNs with $n=24$ are dominant for MPPNs formed from MscS. As observed experimentally \cite{Basta2014} the MPPNs in Fig.~\ref{fig:four} with $n=24$ have the symmetry of a snub cube with MscS located at the polyhedron vertices. We note that MPPNs with $n=12$ and icosahedral symmetry, which exhibit the closest packing of MscS for $10 \leq n \leq 80$ [Fig.~\ref{fig:two}(inset)], yield $\phi(12) \ll 1$ in Fig.~\ref{fig:four} due to the relatively small \cite{Auth2009} $\alpha$ of MscS \cite{Bass2002,Steinbacher2007,Phillips2009}, which results in a large $E_b$ for $n=12$. Our mean-field model of MPPNs predicts, with all parameters fixed directly by experiments \cite{Basta2014,Bass2002,Steinbacher2007,Phillips2009,Rawicz2000}, that MPPNs with $n=24$ have a bilayer midplane radius $R\approx 10$~nm for MPPNs formed from MscS \cite{Basta2014,Wu2013}. Adjusting for the length of the MscS cytoplasmic region $\approx 10$~nm \cite{Steinbacher2007,sm} (Fig.~\ref{fig:one}), the size of the dominant MPPNs predicted by our model is in quantitative agreement with the total MPPN radius $\approx 20$~nm measured in experiments \cite{Basta2014} for $n=24$. Apart from the dominant MPPNs with $n=24$, experiments also suggest \cite{Basta2014} that lipids and MscS can self-assemble into MPPNs with a smaller average radius, but the symmetry of these MPPNs is unclear. The observed sub-dominant MPPNs \cite{Basta2014} may correspond to the low-symmetry structures competing with MPPNs with $n=24$
in Fig.~\ref{fig:four}. In particular, Fig.~\ref{fig:four} predicts that the most abundant low-symmetry MPPNs correspond to MPPNs with $n=20$, $\text{D}_{\text{3h}}$ symmetry, and a radius at the bilayer midplane which is reduced by $\approx 1$~nm compared to MPPNs with $n=24$.

\begin{figure}[t!]
\centerline{\includegraphics[width=\columnwidth]{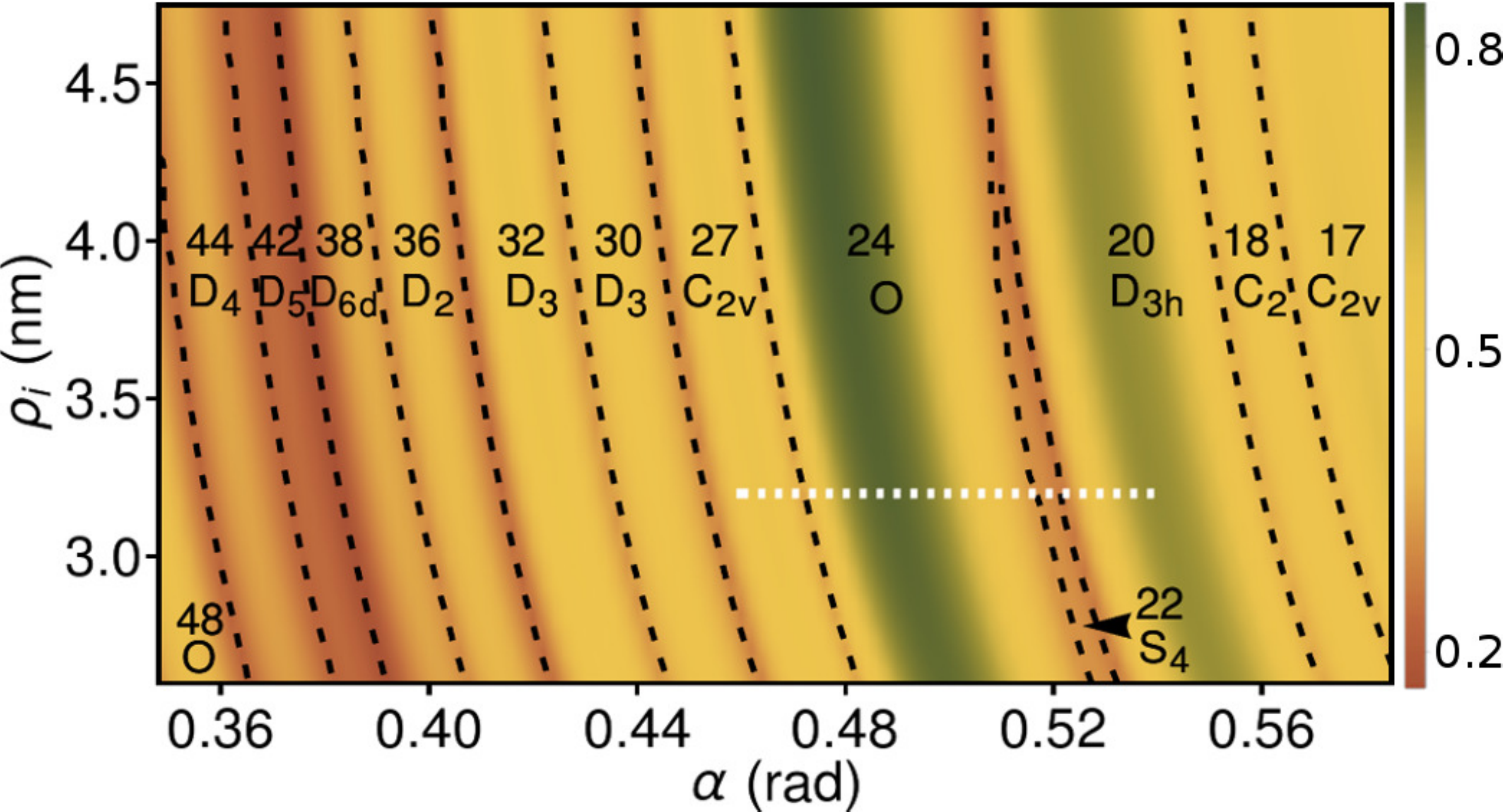}}
\caption{\label{fig:five} (color online). MPPN self-assembly phase diagram
obtained from Eq.~(\ref{eq:nine}) as a function of protein radius $\rho_i$ and bilayer-protein contact angle $\alpha$. The white dashed line indicates the protein radius $\rho_i\approx 3.2$~nm and $\alpha$-range associated with MscS \cite{Steinbacher2007,Phillips2009,sm}. We use the same labeling conventions as in Fig.~\ref{fig:four} with the model parameter values $c\approx7.8\times10^{-8}$ \cite{Basta2014} and $K_b \approx 14$~$k_B T$ \cite{Rawicz2000} corresponding to MPPNs formed from \cite{Basta2014,Wu2013} MscS and diC14:0~lipids.}
\end{figure}

Figure~\ref{fig:four} shows that the dominant MPPN symmetry and size
only weakly depend on $c$. This suggests that even if not all proteins in the system are incorporated into MPPNs, and the effective value of $c$ is smaller than $c\approx7.8\times10^{-8}$ \cite{Basta2014}, the key model predictions discussed above remain unchanged---indeed, smaller $c$ tend to increase the dominance of MPPNs with $n=24$ and $n=20$ (Fig.~\ref{fig:four}) while leaving
the MPPN radius unchanged. In contrast, Fig.~\ref{fig:four} suggests that $\alpha$ is a key parameter setting the preferred symmetry and size of MPPNs. Calculating the MPPN self-assembly phase diagram as a function of $\alpha$ and the protein radius $\rho_i$ (see
Fig.~\ref{fig:five}), we find that the dominant MPPN symmetry is more sensitive to variations in $\alpha$ than $\rho_i$. With the exception of $n=18$, which is almost as closely packed as the locally optimal packing state $n=17$ [Fig.~\ref{fig:two}(inset)], all of the dominant MPPN symmetries in Fig.~\ref{fig:five} correspond to locally optimal protein packings, with $n=24$ yielding the largest $\phi(n)$. Finally, we note that the bilayer bending rigidity $K_b \approx 14$~$k_B T$ \cite{Rawicz2000} of the lipids used for MPPNs \cite{Basta2014} is small compared to other lipids \cite{Rawicz2000,Phillips2009}, and that it has also been suggested \cite{Partenskii2002,Kim2012,Lee2013} that $K_b$ may be increased in the vicinity of membrane proteins. We find \cite{sm} that, as $K_b$ is being increased, the dominance of MPPNs with $n=24$ becomes increasingly pronounced for MPPNs formed from MscS \cite{Basta2014,Wu2013}.

\textit{Conclusion.}\textbf{---}To aid the utilization of MPPNs for high-resolution structural studies \cite{Basta2014,Wu2013} we have developed a simple physical description of MPPNs which connects the symmetry and size of MPPNs to key molecular properties of the lipids and proteins forming MPPNs. Our model accounts for the energy cost of protein-induced lipid bilayer bending deformations \cite{Gozdz2001,Auth2009,Muller2010} and topological defects in protein packing
in MPPNs \cite{Bruinsma2003,Zandi2004}, and the statistical thermodynamics \cite{Bruinsma2003,Zandi2004,Safranbook,BenShaul1994} of MPPN self-assembly. With all model parameters determined directly from experiments, our model correctly predicts the observed \cite{Basta2014} symmetry and size of MPPNs formed from MscS. Our results suggest that the MPPN bending and defect energies determine a lower cutoff on the number of proteins per MPPN, with the MPPN defect energy and thermal effects yielding MPPNs with locally optimal protein packings close to this cutoff as the dominant MPPN symmetry and size. Our model suggests how, through suitable choices of key lipid and protein properties, a range of well-defined MPPN symmetries and sizes can be produced in experiments.

\begin{acknowledgments}

This work was supported by NSF award numbers DMR-1554716 and DMR-1206332, an Alfred P. Sloan Research Fellowship in Physics, the James H. Zumberge Faculty Research and Innovation Fund at the University of Southern California, and the USC Center for High-Performance Computing. We also acknowledge support through the Kavli Institute for Theoretical Physics, Santa Barbara, via NSF award number PHY-1125915. We thank W.~S. Klug, R. Phillips, D.~C. Rees, M.~H.~B. Stowell, and H. Yin for helpful comments.

\end{acknowledgments}

% % \clearpage
% \foreach \x in {1,...,7}
% {%
% % \clearpage
% \includepdf[pages={\x}]{supp}
% }

% \includepdf[pages=-]{supp.pdf}

\end{document}